\documentclass[aps,amsfonts,prl,twocolumn,showpacs]{revtex4}

\usepackage{subfigure}
\usepackage{wrapfig}
\usepackage{textcomp}
\usepackage{graphicx}
\usepackage{amssymb}
\usepackage{amsmath}
\usepackage{bm}
\usepackage{amsmath, amsthm, amssymb}

\def\beq{\begin{equation}}
\def\eeq{\end{equation}}
\def\bea{\begin{eqnarray}}
\def\eea{\end{eqnarray}}
\begin{document}

%\title{Two-dimensional melting in the presence of fractional excitations}

\title{Disclination classes, fractional excitations, and the melting of quantum liquid crystals}

\author{Sarang Gopalakrishnan$^1$, Jeffrey C.Y. Teo$^2$ and Taylor L. Hughes$^2$}

\affiliation{$^1$Department of Physics, Harvard University, Cambridge MA 02138, USA \\ $^2$Department of Physics, University of Illinois at Urbana-Champaign, Urbana IL 61801, USA}

 \date{Feb 14, 2013} 

\begin{abstract}
We consider how fractional excitations bound to a dislocation evolve as the dislocation is separated into a pair of disclinations. We show that some dislocation-bound excitations (such as Majorana modes and half-quantum vortices) are possible only if the elementary dislocation consists of two inequivalent disclinations, as is the case for stripes or square lattices but not for triangular lattices. The existence of multiple inequivalent disclination classes governs the two-dimensional melting of quantum liquid crystals (i.e., nematics and hexatics), determining whether superfluidity and orientational order can simultaneously vanish at a continuous transition.

\end{abstract}

\maketitle

Spatial order (i.e., crystallinity) is found intertwined with exotic quantum-mechanical order in various systems: for instance, with unconventional superfluidity, in the Fulde-Ferrell-Larkin-Ovchinnikov (FFLO) state and related states~\cite{ff, lo, radz, radz2, berg09, agterberg08, agterberg11}; and with topological electronic structure, in weak and crystalline topological insulators/superconductors~\cite{hasan:kane, teo08, fu11, hughes11, turner10, andrei12, andrei13}. In such cases, the defects of the \emph{translational} crystalline order---i.e., \textit{dislocations}---can bind excitations or defects of the intertwined quantum-mechanical order, such as Majorana bound states, protected helical edge modes, or half-quantum vortices~\cite{ashvin, TeoKane10}. More generally, such composite defects, and their proliferation, are central themes in the theory of deconfined quantum criticality~\cite{senthil}. A common feature of the dislocation-bound states noted above is that they are ``$Z_2$'' in character, i.e., two of them combine to give a conventional bound state which is trivial in some sense. 

\begin{figure}[b]
	\centering
	\includegraphics[width=2.5in]{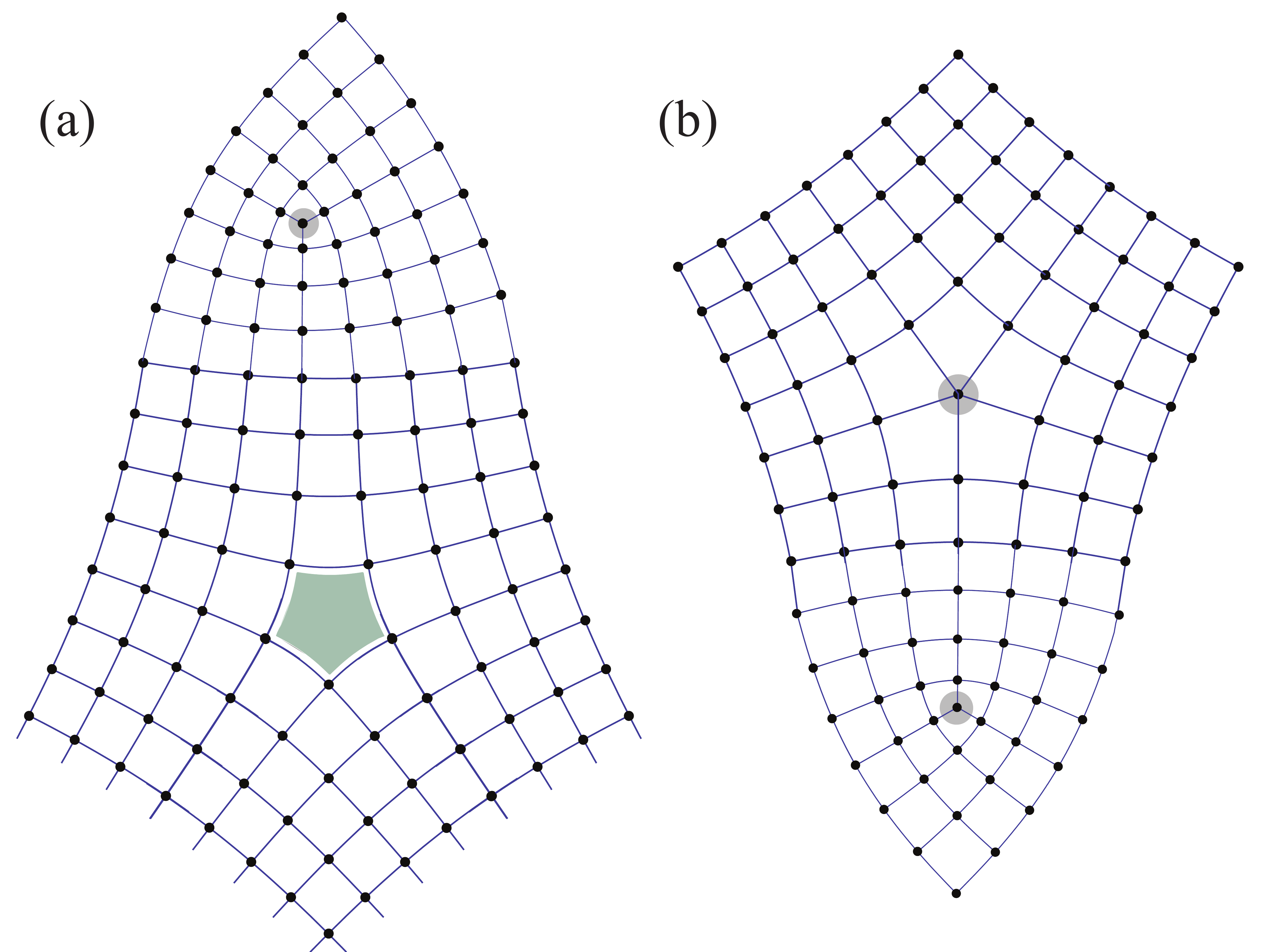}
 \caption{(a), (b) $\pm\pi/2$-disclination dipoles that fuse into dislocations with odd, even Burgers' vectors respectively. }\label{dislocationsquare}
\end{figure}

The fact that dislocations can bind such $Z_2$ states raises a conceptual puzzle because dislocations are themselves composite defects: a dislocation is a bound state of two \textit{disclinations}, which are topological defects of \emph{orientational} crystalline order~\cite{stewart, hn:prl, hn:prb, young79, mermin:review}. When a dislocation is split into two widely separated disclinations (Fig.~\ref{dislocationsquare}), there is \textit{prima facie} no straightforward way for the accompanying bound state to divide or to stretch between the two disclinations; nor is there an obvious principle for assigning the bound state to either disclination. In this work we immediately present the resolution to this puzzle by explaining why the bound state attaches itself to one of the two disclinations, and then we go on to discuss the non-trivial consequences of this concept. 

Following the work of Ref.~\cite{teo:hughes}, we note that certain lattices (including stripes and square lattices) have topologically inequivalent classes of \emph{disclinations}, some of which support exotic bound states on their own. In fact, a dislocation carrying a $Z_2$ bound state is \emph{always} composed of one trivial and one nontrivial disclination and this composition determines how the bound state is resolved when a dislocation splits: it attaches itself to the non-trivial disclination. An immediate corollary is that in lattices without multiple, inequivalent types of disclinations, there will be no non-trivial $Z_2$ bound states on dislocations either. For the case where translational order is intertwined with superfluidity, we show that the presence, or absence, of multiple disclination classes modifies the structure of the phase diagram in two dimensions. For example, in lattices that have only one class of elementary disclination (e.g., hexagonal lattices~\cite{ruegg}), superfluid and liquid-crystalline order can generically vanish at once through a continuous transition; in lattices with inequivalent disclinations, however, this is forbidden except possibly at a multicritical point. While this work only explicitly considers superfluid order, the arguments and main result will also apply to cases where dislocations carry other, perhaps magnetic, defects (as in Refs.~\cite{park:huse, korshunov}). 

The predictions of our analysis are straightforward to test in ultra-cold atomic experiments with FFLO liquid crystals or with spin-orbit coupled condensates~\cite{spielman, campbell, zhai, zhai2, mondragon, galitski, sg:lamacraft}. Furthermore, our central result regarding inequivalent disclinations has immediate experimental consequences for weak topological insulators and superconductors where our arguments imply that it is the disclination core-size, rather than the dislocation core-size, that sets the confinement scale of bound states, hybridization energy gaps, etc. This prediction is particularly salient to experiments, as the conditions under which dislocations are common and easy to probe (i.e., those under which crystalline order is weak) are precisely those under which the dislocations consist of two weakly bound disclinations.

\begin{figure}[t!]
	\centering
		\includegraphics{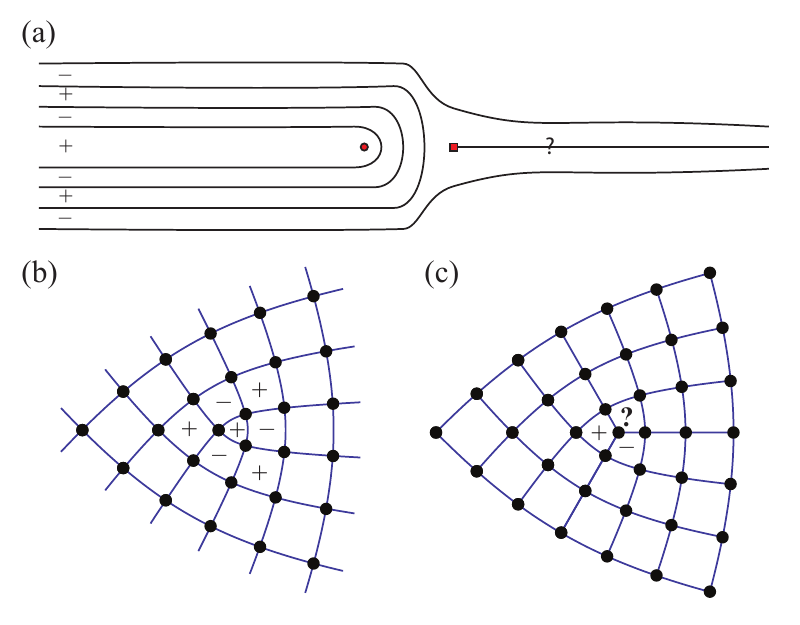}
	\caption{(a)~Two inequivalent classes of disclinations of a two-dimensional striped superfluid. Lines denote nodes of the order parameter; the sign of the condensate between these is indicated on the left. The crest-centered disclination (red circle) involves no sign-frustration, whereas the trough-centered disclination (red square) does involve sign-frustration. (b), (c)~Examples of the two disclination classes for a square-lattice striped superfluid, indicating the absence and presence of sign-frustration respectively.}
	\label{fig:disclins}
\end{figure}

%%%%%%Revised paragraph begins%%%%%%%%%%%%%
\textit{General classification of disclinations}. 
We begin by reviewing the results of Ref.~\cite{teo:hughes}; we focus on three cases, namely stripes, square lattices, and triangular lattices. A crucial observation in the analysis of Ref.~\cite{teo:hughes} is that the holonomy around a disclination can take multiple discrete values depending on where the disclination is centered. The holonomy $h$ is simply the element in the planar space group accumulated along a loop in the lattice around the disclination. It contains a rotation component (Frank angle) and translation component (Burgers' vector)~\cite{stewart}. The group element is path independent up to conjugation $h\mapsto ThT^{-1}$ by a translation $T$ of the starting point of the loop. Disclinations are therefore topologically classified by conjugacy classes of the planar space group; these classes correspond to point-group rotations with inequivalent rotation centers. Thus, stripes have two classes of $\pi$ disclinations, namely disclinations centered on a stripe or halfway between two stripes [Fig.~\ref{fig:disclins}(a)], and an elementary dislocation contains one disclination of each class. For square lattices, there is a similar $Z_2$ classification of $\pi/2$ disclinations [Fig.~\ref{fig:disclins}(b),(c)], which are either vertex-centered or plaquette-centered, and again an elementary dislocation contains one disclination of each class. However, for a triangular lattice there is only one class of $\pi/3$-disclinations and $2\pi/3$-disclinations are $Z_3$-classified. One can easily check that, because 2 and 3 are coprime, none of the three types of $2\pi/3$ disclinations carries a $Z_2$ bound state, and it follows that all dislocations are topologically trivial. 
%%%%%%Revised paragraph ends%%%%%%%%%%%%%

\textit{Striped superfluids}. We now address the nature of disclination-vortex bound states, beginning with the case of a striped superfluid that is described by a Bose condensate of the form $\psi(\mathbf{x}) = \psi \cos(\mathbf{k \cdot x} + \phi) \exp(i\theta)$, where $\psi$ is taken to be a fixed amplitude and $\theta$ and $\phi$ are phases associated with the superfluid and translational Goldstone modes respectively. The canonical example of such a state is the Fulde-Ferrell-Larkin-Ovchinnikov state proposed for spin-polarized superconductors~\cite{ff, lo, radz, radz2, hulet1, hulet2}; a closely related state, which may occur in layered cuprate superconductors, is the pair-density-wave state~\cite{berg09}. Furthermore, ultracold atomic gases with spin-orbit coupling offer a purely bosonic analog of these states~\cite{mondragon, zhai, zhai2, sg:lamacraft, barnett}. For specificity, we address the rotationally invariant case; thus, our analysis directly applies to FFLO states and to Bose gases subject to a Rashba spin-orbit coupling~\cite{campbell, zhai, zhai2, sg:lamacraft}. As we are primarily concerned with topological features rather than energetics we could restore weak anisotropy without affecting our conclusions.

The striped superfluid has three easily identifiable types of topological defects~\cite{berg09}: (i)~ a pure dislocation, in which $\phi$ winds by $2\pi$, (ii)~a pure vortex, in which $\theta$ winds by $2\pi$, and (iii)~a half-vortex-dislocation bound state, in which both $\phi$ and $\theta$ wind by $\pi$. Since defects of type (i) and (ii) can be written as dipoles of (iii), it suffices to understand how the vortex-dislocation bound state separates into disclinations. The $Z_2$ classification of disclinations for stripes given above provides a direct (and hitherto undiscussed) geometric criterion for understanding how this separation occurs. We know that the half-quantum vortex bound to a dislocation compensates for the change in sign of the order parameter going around the dislocation. Of the two kinds of possible disclinations, only the node-centered disclination (type-A) involves a frustration of the order-parameter sign. Therefore, as an elementary vortex-dislocation bound state consists of one node-centered and one anti-node centered (type B) disclination, the type-A disclination inherits the half-vortex when the dislocation unbinds.

 We now turn to the melting of a striped superfluid. As is well known~\cite{radz}, in a rotation-invariant two-dimensional system, dislocations proliferate at any nonzero temperature, reducing the smectic to a nematic. While  cold atom systems are never exactly rotation symmetric, the temperature scale set by the anisotropy (at which dislocations proliferate) can be taken to be much lower than that at which the nematic melts. In what follows, therefore, we consider this nematic phase. From above we know a type-A disclination carries both a Frank angle (an orientational ``charge'') and a vorticity (a superfluid charge), whereas a type-B disclination  only has an orientational charge. In what follows we denote the elementary disclinations and their composites using the notation $(\alpha, \beta)$ where $\alpha$ is the orientational charge and $\beta$ is the superfluid charge. In this notation, a pure dislocation is $(0,0)$ (and can be constructed as a dipole of two type-A or type-B disclinations with opposite $\alpha$); a pure vortex has charge $(0,2)$ and can be composed of two type-A disclinations of opposite $\alpha$ (Frank angle); and finally, a half-vortex half-dislocation bound state consists of one type-A and one type-B disclination, and thus carries charges $(0,1)$. 

The low-energy theory of a nematic superfluid is given, in the one-constant approximation~\cite{stewart}, by the expression

\beq
\mathcal{H}_{el.} = \kappa \left[ (\nabla\mathbf{ \cdot n})^2 + (\nabla \mathbf{ \times n})^2 \right] + \rho_s |\nabla \theta|^2,
\eeq
where $\mathbf{n}$ is a unit vector normal to the local orientation of the nematic, $\kappa$ is the Frank constant governing orientational stiffness, and $\rho_s$ is the superfluid stiffness. This elastic theory can be analyzed using standard Coulomb gas methods~\cite{berg09, minnhagen}; this involves transforming the elastic theory to a theory in terms of the topological defects discussed above:

%%%%(alpha,beta)<------>(beta,alpha)
\bea
\mathcal{H}_{c} & = & \frac{T}{2\kappa} |\nabla \phi_n|^2 + \frac{T}{2\rho_s} |\nabla \phi_s|^2\\
& & - \sum_{\alpha\beta} g_{\alpha,\beta} \cos(\pi \alpha \phi_n) \cos(\pi \beta \phi_s). \nonumber
\eea
One can now compute the temperature at which a given defect unbinds by computing the scaling dimensions of the $g_{\alpha,\beta}$ of every topological defect $(\alpha,\beta)$ around the noninteracting fixed point defined by the quadratic terms. These scaling dimensions are generically given by the expression, $\Delta_{\alpha,\beta} \equiv (\pi/T) [\alpha^2 \kappa + \beta^2 \rho_s]$~\cite{berg09}; a defect with indices $(\alpha,\beta)$ proliferates when $\Delta_{\alpha,\beta} = 2$. 

\begin{figure}[t!]
	\centering
		\includegraphics{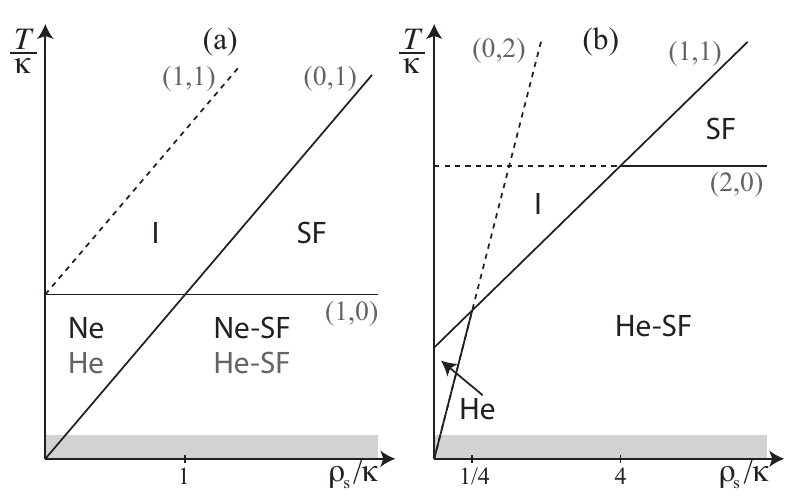}
	\caption{(a) Phase diagram of a striped, or square-lattice, superfluid in two dimensions, for a fixed Frank constant, as a function of superfluid stiffness and temperature. Solid lines denote phase transitions, due to the proliferation of defects (indexed by their orientational and superfluid charges); the dashed line represents the temperature at which type-A disclinations would (hypothetically) have proliferated if the system were still ordered. The four phases are nematic/hexatic (Ne/He), nematic/hexatic superfluid (Ne/He-SF), isotropic superfluid (SF), and isotropic normal (I). (b) Phase diagram of a triangular-lattice superfluid [case (a)]. The notation is the same as that in (a); note that a direct transition from the hexatic superfluid to the isotropic normal state, absent in (a), appears here.}
	\label{fig:phasediag}
\end{figure}

In the nematic superfluid, all charge is confined, so that the basic components of the equilibrium state are neutral dipoles of two type-A or two type-B disclinations. The type-A disclinations are bound by both superfluid and elastic charge; thus, they are more tightly bound than the type-B disclinations, in this phase, and can never proliferate before the type-B disclinations. Thus, from our analysis  we find that there is generically no direct transition from the nematic superfluid to an isotropic normal phase contrary to  Ref.~\cite{barci}. In the Coulomb gas language, one can straightforwardly see this as follows. Depending on the values of $\rho_s/\kappa$, the excitations with the lowest threshold for unbinding are type-B disclinations [$(1,0)$] or orientationally neutral half-vortex half-dislocations [$(0,1)$]; 
other defects are always less relevant than these. For small superfluid densities, the half-vortex half-dislocation is the first to proliferate, leading to a nematic but non-superfluid state. For larger superfluid densities, the type-B disclination is the first to proliferate, giving rise to an isotropic superfluid state. The resulting phase diagram is shown in Fig.~\ref{fig:phasediag}(a). At this level of analysis the two transitions meet at a multicritical point, at $\kappa = \rho_s$, which is the only point at which a direct transition out of the nematic superfluid into a normal phase is possible; however, nonlinearities not included in this analysis might change the phase diagram in the vicinity of the multicritical point. (Note that, in experiments with spin-orbit coupled Bose gases, one can tune the ratio $\kappa/\rho_s$ by tuning the spin-dependent scattering lengths: $\kappa/\rho_s \ll 1$ near the stripe-to-plane-wave phase transition~\cite{zhai, zhai2, mondragon, galitski, sg:lamacraft}, while the opposite limit holds deep in the striped phase.) 

 We now extend our analysis to the case of hexagonal and square-lattice pair-density waves with sign-changing order parameters. These can arise due to nonlinear interactions between stripes, or can be directly engineered by tuning interactions (e.g., cavity-mediated~\cite{sg:np, sg:pra} or fermion-mediated interactions)  to be peaked at certain momenta. A specific case of  a pair-density-wave state on a hexagonal lattice was recently discussed in Ref.~\cite{agterberg11} with an analysis that ignored the effects of disclinations. However, disclinations play a crucial role in the melting transition and we find, remarkably, that the nature of two-dimensional melting depends strongly on the lattice geometry, via the classification of disclinations. 
 
Unlike striped states, crystalline states can sustain quasi-long-range order at finite temperatures in two dimensions; however, they generally melt in two steps, with the dislocations proliferating first and then the disclinations~\cite{hn:prl, hn:prb, young79, mathy}. We shall analyze the intermediate ``hexatic'' phase, in which pure dislocations have proliferated 
(so that translational order is short-range) but disclinations are still confined. For square lattices (i.e., checkerboard arrangements of the order parameter), as with stripes, there are two inequivalent classes of disclinations;  as can be seen in Fig.~\ref{fig:disclins}, only one of the two classes has sign-frustration and thus binds a half-vortex. Thus, the overall phase diagram in the square-lattice case is exactly as in the striped case: the first defect to proliferate is either the neutral disclination or the vortex-dislocation bound state, and in either case the transition out of the hexatic superfluid is into either a hexatic or a superfluid state.

\begin{figure}[t!]
	\centering
		\includegraphics{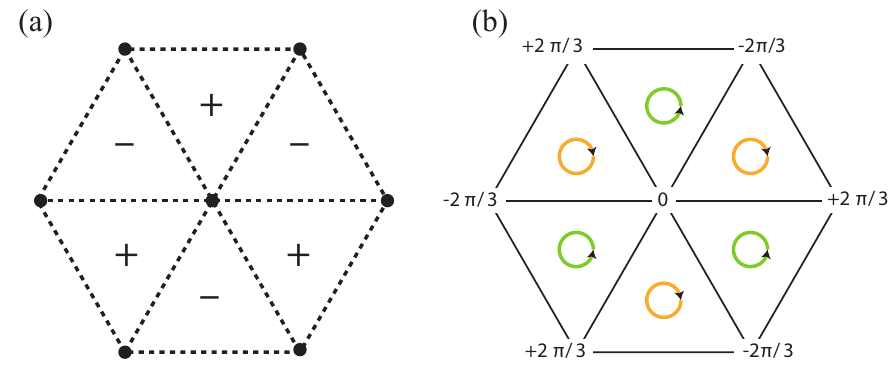}
	\caption{Two possible varieties of hexagonal-lattice superfluid. (a)~Hexagonal lattice of nodes of the order parameter, corresponding to a honeycomb lattice of the particle density, $+,-$ represent the sign of Bose condensate $\psi$; (b)~Hexagonal lattice in the particle density, corresponding to a honeycomb lattice of vortices, $0,\pm2\pi/3$ are phases of the condensate $\psi$.}
	\label{fig:hexlat}
\end{figure}

\begin{figure}[t!]
	\centering
	\includegraphics[width=3.5in]{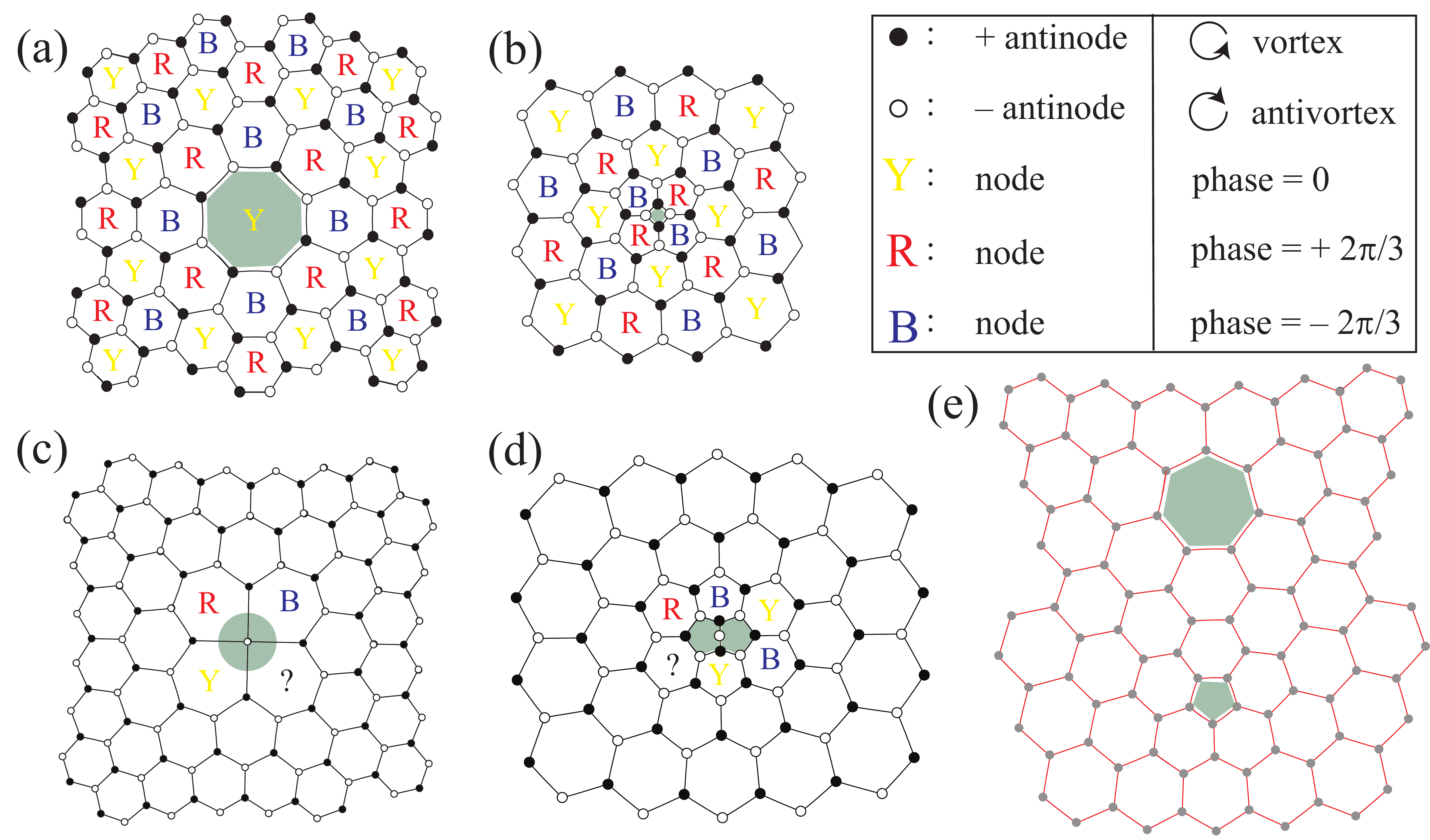}
 \caption{(a-d) Inequivalent trivial $\pm2\pi/3$-disclinations. None breaks the $\bullet,\circ$-sublattice order (no nearest vertices are of same type). Tri-coloring ordering (no adjacent plaquettes have same color) is preserved for (a) and (b) and is broken for (c) and (d). (e) Topological $\pm\pi/3$-disclination dipole. Each breaks $\bullet,\circ$-sublattice and tri-coloring order.}\label{disclinationhoneycomb}
\end{figure}

%%%%%%%%%%%%Revised paragraph (added explanation to the honeycomb figure)%%%%%%%%%%%%
We now turn to triangular/hexagonal lattices where we must distinguish between two types of order parameter configurations: (a)~that in which the nodes of the order parameter form a triangular lattice [Fig.~\ref{fig:hexlat}(a)], and the antinodes occupy the dual honeycomb lattice; and (b)~that in which the antinodes form a triangular lattice [Fig.~\ref{fig:hexlat}(b)], and the nodes occupy the dual honeycomb lattice. This complication does not appear in the square lattice because it is self-dual. We consider case~(a) first. We see the $\pm\pi/3$ disclinations on the triangular lattice in Fig.~\ref{disclinationhoneycomb}(e) break the $\bullet,\circ$-sublattice order that corresponds to the opposite signs in Fig.~\ref{fig:hexlat}(a) and therefore invariably carries a half-vortex, i.e.  it carries charges $(1,1)$. These elementary disclinations can be combined to form pure vortices $(0,2)$ or pure double disclinations $(2,0)$ equivalent to the $2\pi/3$ disclinations in Fig~\ref{disclinationhoneycomb}(a-d). Depending on $\rho_s/\kappa$, any of the three defects might proliferate first. In particular, the proliferation of elementary $(1,1)$ disclination-vortex states, which is favored when the orientational and superfluid stiffness are comparable [i.e., $1/4 \leq \kappa/\rho_s \leq 4$], can give rise to a direct Kosterlitz-Thouless transition between the nematic superfluid and an isotropic normal state, as indicated in Fig.~\ref{fig:phasediag}(b). (The formal structure of the Coulomb-gas problem, and thus the overall structure of the phase diagram, are then identical to that in Ref.~\cite{berg09}.)
%%%%%%%%%%%%End revised paragraph (added explanation to the honeycomb figure)%%%%%%%%%%%%

%%[sg-revised version]

We now briefly comment on case~(b), in which the antinodes form a hexagonal lattice~\cite{agterberg11}; one can alternatively regard this state as a honeycomb lattice in which the $\bullet$-sites are occupied by vortices and the $\circ$-sites by antivortices in Fig~\ref{disclinationhoneycomb}. Formally, this situation is somewhat different from that discussed so far, as the relative phase between neighboring antinodes is not $\pi$ (i.e., a minus sign) but is instead $2\pi/3$. This modulation of the superfluid phase can be understood as a tri-coloring structure in the honeycomb plaquettes in Fig~\ref{disclinationhoneycomb} where no adjacent plaquettes carry same color. It was argued in Refs.~\cite{agterberg11, korshunov, park:huse} that such a superfluid supports a topological defect with a phase twist of $2\pi/3$ (i.e., a vorticity of $1/3$) bound to an elementary dislocation. In the hexatic phase, dislocations are screened; hence this defect carries only the elementary superfluid charge (i.e., it is simply an elementary vortex). When the orientational stiffness exceeds the superfluid stiffness, this proliferates; in the opposite limit, the first defects to proliferate are the plaquette-centered $2\pi/3$ disclinations [Fig.~\ref{disclinationhoneycomb}(a,b)], which do not violate the tri-coloring pattern, and are therefore neutral with respect to superfluidity. Thus the overall phase diagram has the form shown in Fig.~\ref{fig:phasediag}(a). Note that such a phase diagram could arise on the triangular lattice precisely because the superfluid defect was $Z_3$, rather than $Z_2$, in character. A nontrivial $Z_3$ defect can bind to a dislocation \textit{even if} the dislocation consists of two equivalent disclinations, because $2 = 1 + 1$ and $1 = 2 + 2$ mod 3, so that the original puzzle discussed in the introduction does not apply. Thus, case~(b) of the triangular-lattice superfluid can be regarded as a simple counterexample illustrating the crucial role played in our analysis by the $Z_2$ nature of the exotic bound state.

 Finally, we sketch the implications of our results for weak topological insulators and superconductors. It is well-known that the presence of low-energy bound states on a dislocation is determined by the weak topological index~\cite{ashvin}; as the weak-index is \emph{always} zero for a triangular lattice, it follows that there are no bound states. On the square lattice, one can explicitly see from the construction of Ref.~\cite{teo:hughes} that a weak topological invariant entails the presence of inequivalent disclinations. We note that, although no analogous classification of topological insulators (i.e., not assuming particle-hole symmetry) has yet been performed, one can explicitly see that in the case of a weak topological insulator layered along the $z$ axis, an edge dislocation line along the $z$ axis containing helical edge modes can be decomposed into a disclination line that carries helical modes and one that does not. 
These observations imply that the transverse size of the bound states is set by that of the \emph{disclination} core, which is typically smaller than the dislocation core. The presence of bound states also generates an effective attraction between disclinations of the same topological type; however, one expects this to fall off exponentially with separation and therefore not to have qualitative implications for melting transitions.

In summary,  we have shown that the presence, or absence, of inequivalent disclination classes has a qualitative impact on the phase diagrams of pair-density-wave states: if there are inequivalent disclination classes, the nematic superfluid must melt in two steps; whereas if all disclinations are equivalent, then a direct transition from the nematic superfluid to an isotropic normal state is possible. These predictions are straightforward to test experimentally in ultracold atomic gases, as vortex-unbinding can be seen via interferometry~\cite{hadzibabic} and orientational order via time-of-flight imaging~\cite{bloch:zwerger}. The nature of the interplay between dislocations, disclinations, and exotic bound states can be generalized from superfluidity to e.g. spin-density-wave order as in the condensation of magnons with spin-orbit coupling~\cite{turlakov}. 
Finally, we have argued that the spatial extent of the modes bound to dislocations in weak topological insulators or superconductors is governed by the disclination size (i.e., the inverse orientational stiffness), which is typically smaller than the dislocation size (i.e., the inverse shear stiffness). The extensions of this classification of disclinations to other classes of states, such as nematic Chern insulators~\cite{xiaoliang, xiaoliang2}, and to quantum melting~\cite{zaanen2011, zaanen2013} will be considered in future work.

\textit{Acknowledgments}. S.G. is indebted to Erez Berg, David Huse, and Andreas R\"{u}egg for helpful discussions. This work was supported in part by the Harvard Quantum Optics Center (S.G.),  the Simons Fellowship (J.C.Y.T.), and  by ONR award N0014-12- 1-0935 (T.L.H.).

%\bibliography{disclin}

\end{document}